\documentclass[journal=jpcl,manuscript=letter]{achemso}

\usepackage[version=3]{mhchem} 

\author{Riccardo Capelli}
\affiliation[INM9]{Computational Biomedicine (INM-9/IAS-5), Forschungszentrum J\"ulich, Wilhelm-Johnen-Stra\ss{}e, D-52425 J\"ulich, Germany}
\email{r.capelli@fz-juelich.de}
\author{Paolo Carloni}
\affiliation[INM9]{Computational Biomedicine (INM-9/IAS-5), Forschungszentrum J\"ulich, Wilhelm-Johnen-Stra\ss{}e, D-52425 J\"ulich, Germany}
\alsoaffiliation{Department of Physics, RWTH Aachen University, D-52078 Aachen, Germany}
\author{Michele Parrinello}
\affiliation[ETHZ-USI]{Department of Chemistry and Applied Biosciences, ETH Z\"urich, c/o USI Campus, Via Giuseppe Buffi 13, CH-6900 Lugano, Ticino, Switzerland}
\alsoaffiliation{Facolt\`a di Informatica, Istituto di Scienze Computazionali, Universit\`a della Svizzera Italiana (USI), Via Giuseppe Buffi 13, CH-6900, Lugano, Ticino, Switzerland}
\alsoaffiliation{Istituto Italiano di Tecnologia, Via Morego 30, I-16163 Genova, Italy}

\title{Exhaustive Search of Ligand Binding Pathways via Volume-based Metadynamics}

\abbreviations{MD,ES,CV,WT-MetaD,T4L}

\begin{document}

\begin{tocentry}
\centering
 \includegraphics[]{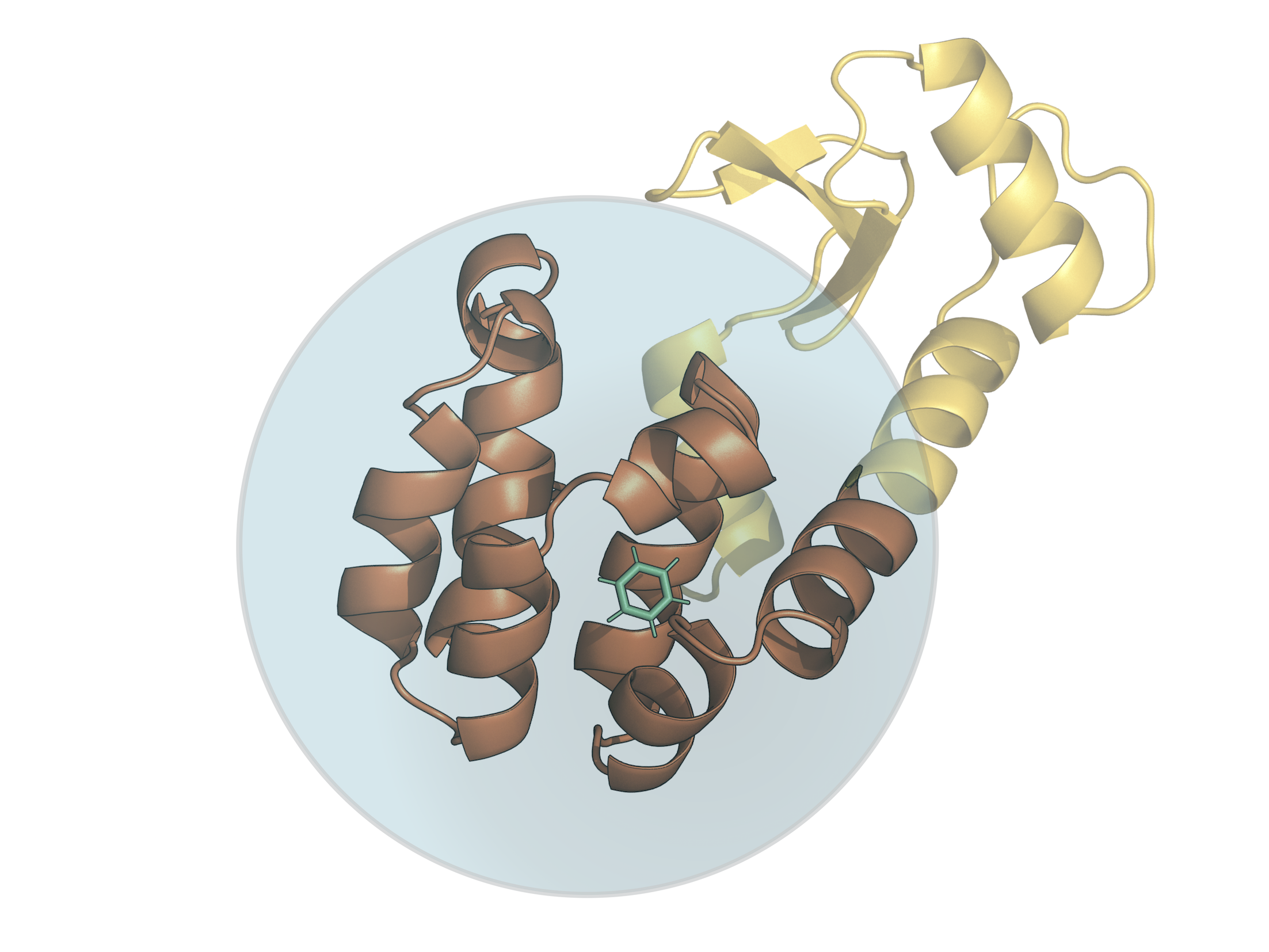}
\end{tocentry}

\begin{abstract}
Determining the complete set of ligands' binding/unbinding pathways is important for drug discovery and to rationally interpret mutation data. Here we have developed a metadynamics-based technique that addressed this issue and allows estimating affinities in the presence of multiple escape pathways. Our approach is shown on a Lysozyme T4 variant in complex with the benzene molecule. The calculated binding free energy is in agreement with experimental data. Remarkably, not only we were able to find all the previously identified ligand binding pathways, but also we uncovered 3 new ones. This results were obtained at a small computational cost, making this approach  valuable for practical applications, such as screening of  small compounds libraries. 
\end{abstract}

\begin{figure}[h!]
 \includegraphics[width=0.7\textwidth]{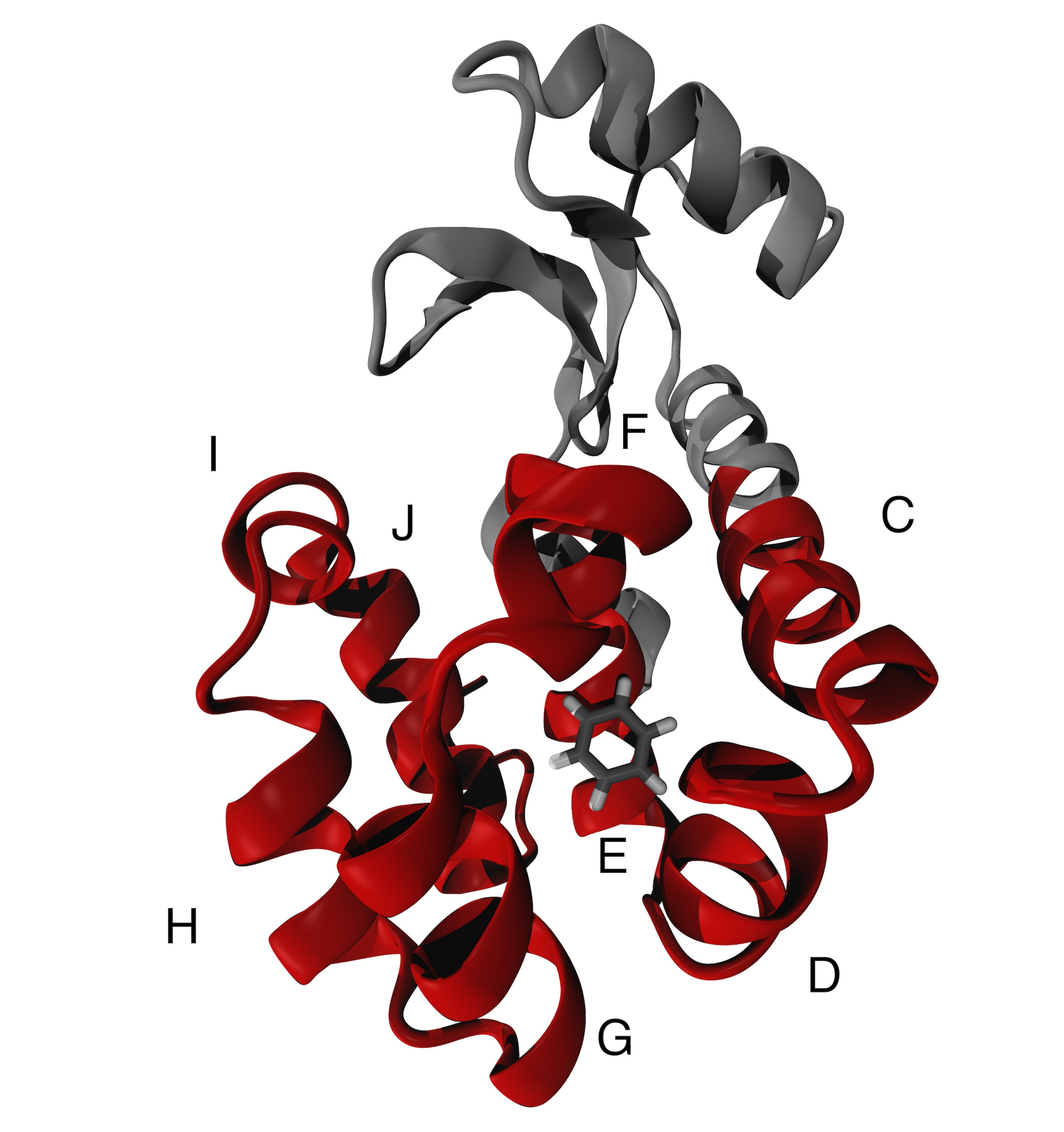}
 \caption{Cartoon representation of the T4L protein. The C-terminal domain is depicted in red color. The benzene ligand is respresented in dark grey  licorice. The N-terminal domain, not involved in the binding process, is in light grey. The helices of the C-terminal domain from C to H are labeled following Rydzewski et al. \cite{rydzewski2018finding}}.
 \label{fgr:T4L}
\end{figure}
\begin{figure}[h!]
 \includegraphics[width=0.7\textwidth]{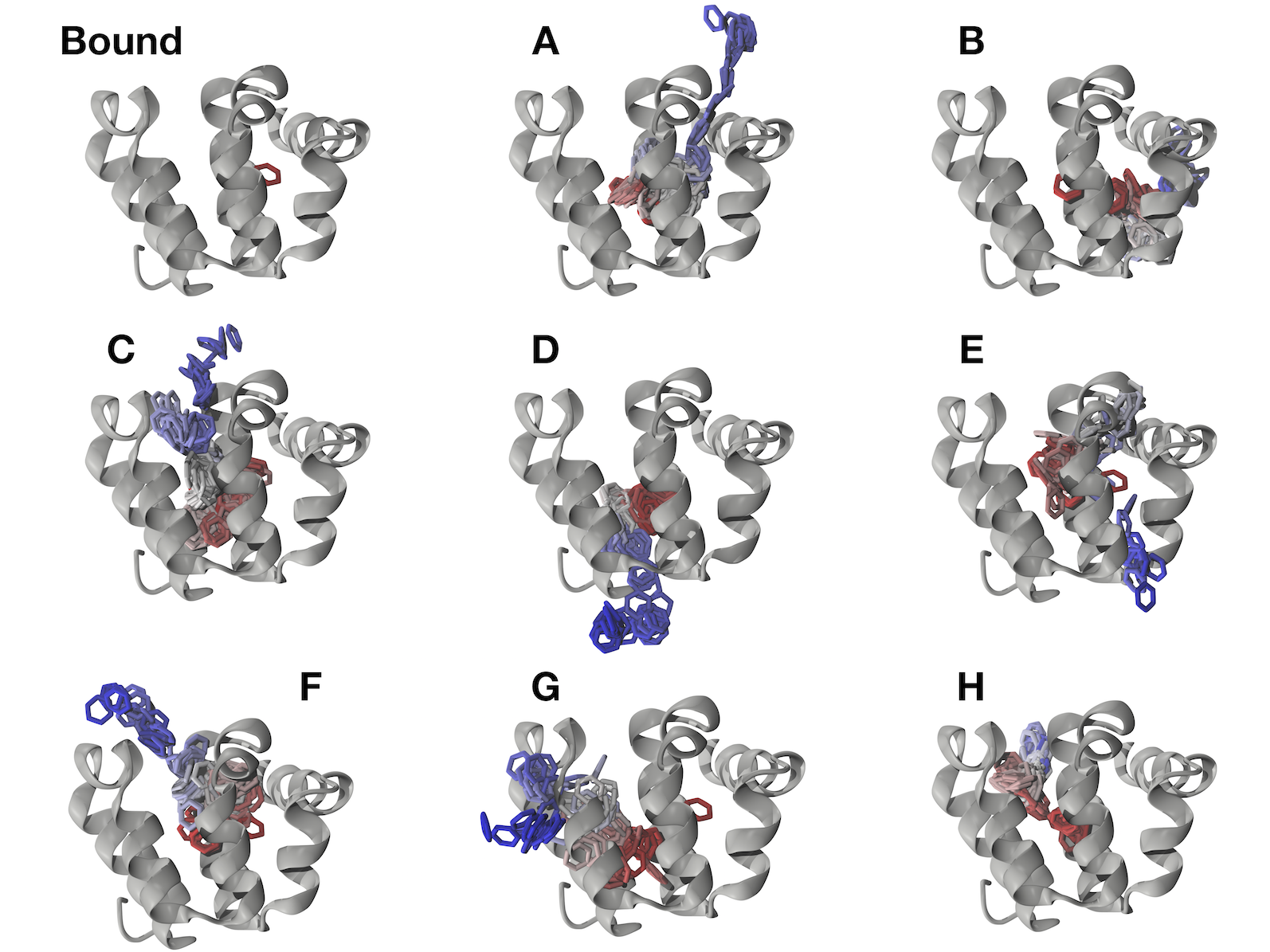}
 \caption{Bound state and binding pathways (\textbf{A} to \textbf{H}) of benzene to T4L C-terminal domain (gray). The benzene ligand is colored from red to blue on passing from the  bound to the unbound state.}
 \label{fgr:pathways}
\end{figure}

Describing mechanisms and energetics of ligand binding and unbinding from their targets is of great importance in drug design. The prediction of poses, affinities, and binding kinetics helps in understanding the effect of chemical decorations on the ligand and/or in mutations in the host system. \cite{held2011mechanisms}. These important problems are most often associated with rare events and they can be adequately studied only with enhanced sampling (ES) methods, that allow estimating the free energy as a function of appropriate collective variables (CVs). In this context, we have recently used one of such methods (funnel\cite{limongelli2013funnel} well-tempered metadynamics\cite{laio2002escaping,barducci2008well}), combined with novel and powerful dimension reduction approach to identify the CVs, to predict the free energy landscape of a ligand binding to a typical membrane receptor, the muscarinic M\textsubscript{2}\cite{capelli2019chasing}. The picture that did emerge from our study was relatively simple, with two rather diverse escape routes of the ligand towards the extracellular region. However, proteins can exhibit a much larger flexibility with many more diverse escape pathways\cite{stank2016protein}. This is the case of the well-studied lysozyme T4 L99A variant (hereafter T4L) \cite{eriksson1992cavity,morton1995specificity,baase2010lessons} (see Figure \ref{fgr:T4L}) in complex with benzene. The protein fold consists of two domains: the N-terminal one (residues 1-70) formed by 3 $\alpha$-helices and 3 antiparallel $\beta$-strands, and a barrel-shaped C-terminal domain (residues 71-162) formed by 8 $\alpha$-helices, in which the ligand accommodates. The protein features several binding/unbinding pathways, whose complete picture is not yet clear; indeed, a variety of ES approaches, with different force fields and solvent representations (see Table \ref{tab:pathways}) have identified up to five pathways involving the C-terminal domain shown in Figure \ref{fgr:pathways} (See SI).

Clearly, a computationally inexpensive methodology which would permit as an exhaustive as possible exploration of all the binding pathways, along with their energetics, would be highly desirable for a theory point of view and for drug discovery applications.

Here, we introduce a new practical method for addressing this issue at a modest computational cost. We shall do this in the frame of Well-tempered Metadynamics (WT-MetaD). As it is well known, WT-MetaD relies on the choice of an appropriate set of CVs. We take advantage of the fact that T4L flexibility is limited and thus its center of mass can be defined once and for all. Centered around this position we consider a sphere of finite radius $\rho_{\text{s}}$, larger than the radius of gyration of our protein. Within this volume we use as CVs the spherical coordinates $(\rho,\theta,\varphi)$. A repulsive potential is added at the border of the sphere so as to limit the volume in the solvated state that needs to be sampled, increasing the probability of a subsequent recrossing event. The restraining potential is in the form
\begin{equation}
\label{eqn:restraint}
U_{\text{s}}(\rho(t)) = 
\begin{cases}
\frac{1}{2}k \left(\rho(t) - \rho_{\text{s}}\right)^{2} ~~~ \text{if } \rho(t) > \rho_{\text{s}} \\
0 ~~~~~~~~~~~~~~~~~~~~~~~\text{otherwise}
\end{cases}
\end{equation}
where $k$ has to be large enough to prevent the ligand escaping from the confining volume. In equation (\ref{eqn:restraint}) $\rho(t)$ is the distance of the ligand from the center of mass of the target protein, and $\rho_{\text{s}}$ is the radius of the spherical restraint. \\
This approach is an extension of funnel metadynamics, where a funnel-shaped potential limits the volume accessible to the ligand in the solvated space, considering a single port of entry for the ligand. Here, we are instead able to observe any binding/unbinding pathway accessible. 

The application of such a restraining potential causes a change in the translational entropy of the solvated states. With the same spirit of previous works\cite{allen2004energetics,limongelli2013funnel}, the correction to the binding free energy due to this constraint can be estimated to be
\begin{equation}
\label{eqn:correction}
\Delta G_{b}^{0} = -RT \log\left( C^{0} K_{b} \right)=\Delta G_{\text{MetaD}} - RT \log\left( \frac{V^{0}}{\frac{4}{3} \pi \rho_{\text{s}}^{3} -V_{\text{prot}}}\right)
\end{equation}
where $R$ is the gas constant, $T$ is the system temperature, $K_{b}$ is the binding constant, $C^{0} = 1660 \text{\AA}^{3}$ is the standard concentration, $V^{0}$ is its reciprocal, $\Delta G_{\text{MetaD}}$ is the binding free energy obtained by WT-MetaD, and $V_{\text{prot}}$ is the volume of the protein inside the restraining potential. The derivation of this correction can be found on the SI. \\ 

\begin{figure}[h!]
 \includegraphics[width=0.3\textwidth]{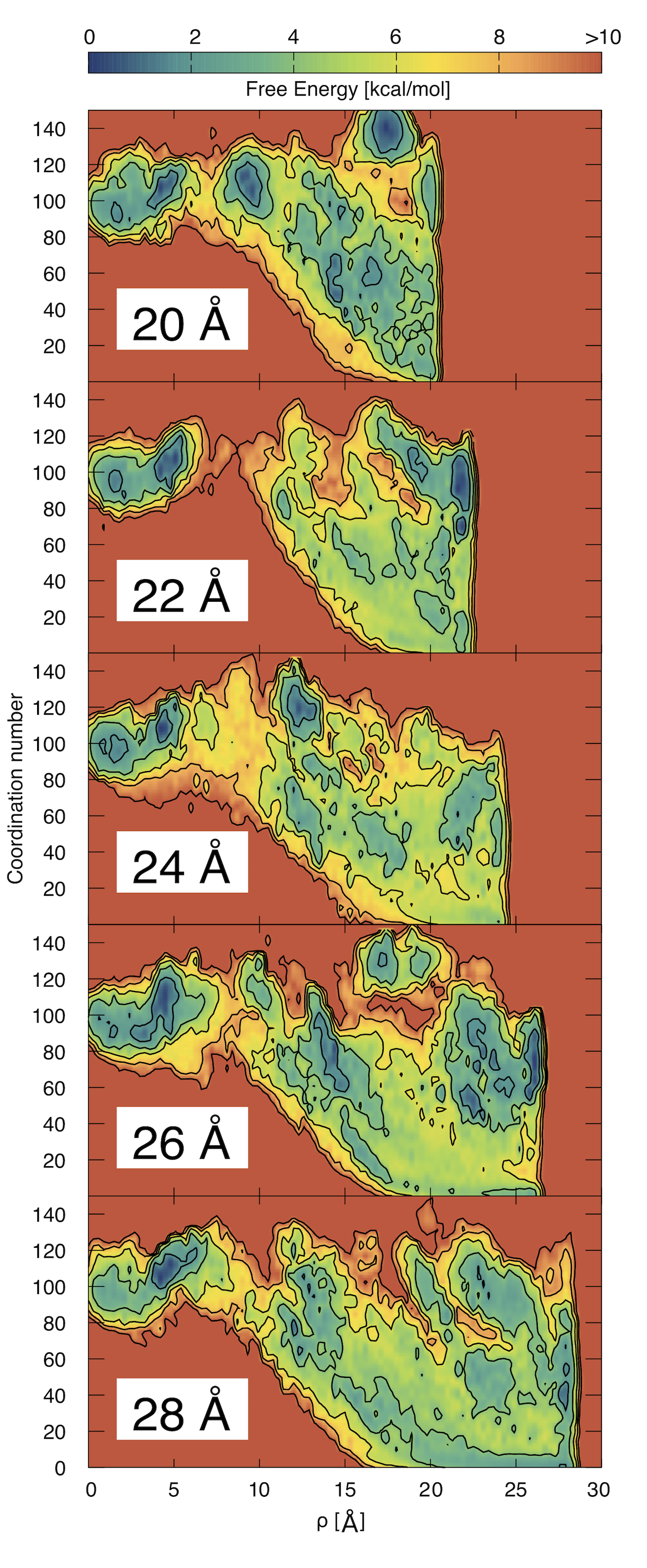}
 \caption{Free energy surfaces in function of the restraint radius $\rho_{\text{s}}$.}
 \label{fgr:FES}
\end{figure}

To build our model, we used an experimental X-ray structure of T4L /benzene complex  (PDB code: 1L84\cite{eriksson1992cavity}). Details on the system preparation, equilibration and run parameters can be found in the SI. The simulation were performed with GROMACS 2018.3\cite{abraham2015gromacs}, patched with Plumed 2.5\cite{tribello2014plumed}. \\
We performed 5 different 200-ns long WT-MetaD simulations, using different values of $\rho_{\text{s}}$. After reaching convergence, the calculated free energy surface (FES) as a function of $\rho, \theta, \varphi$ is projected (with a reweighting procedure\cite{tiwary2014time}) on two more informative CVs. These are the distance from the center of mass of the C-terminal domain $\rho$ and a CV that measures whether the ligand is in contact with the host protein, measured by the coordination number $C_N$
\begin{equation}
  C_{N} = \sum_{i \in A} \sum_{j \in B} \frac{ 1 - \left(\frac{{\bf r}_{ij}}{r_0}\right)^n } { 1 - \left(\frac{{\bf r}_{ij}}{r_0}\right)^m }
\end{equation}
where A is the set of non-hydrogen  atoms of the ligand, B is the set of non-hydrogen atoms of the protein, $r_{0}$=4.5 \AA{ },is the threshold to define a formed contact, $r_{ij}$ is the distance between  atoms $i$ and $j$, and $m=6$ and $n=12$ are the exponents of the switching function. \\
The projected FES (Figure \ref{fgr:FES}) is shown as a function of different $\rho_{\text{s}}$ in order to study the influence of this parameter on the results.
It is evident that, especially for shorter values of $\rho_{\text{s}}$, the presence of the restraining potential induces some artificial minima at the border of the ligand accessible volume. However, these artifacts decrease with increasing the $\rho_{\text{s}}$ value and in our case the estimate of $\Delta G^{0}_{b}$ does not change in a sensible way with $\rho_{\text{s}}$, appearing to be consistent with the experimental data (see Table \ref{tab:deltaG}).

\begin{table}
\caption{Absolute free energy differences obtained with our technique at different radii of the restraining sphere compared with experimental data.}
 \label{tab:deltaG}
 \begin{tabular}{cccc}
  \hline
  $\rho_{\text{s}}$ [\AA] & $\Delta G_{\text{MetaD}}$ [kcal/mol] & $T\Delta S$ [kcal/mol] & $\Delta G^{0}_{b}$ [kcal/mol] \\
  \hline
  20 		& $-3.6 \pm 0.5$	& $1.6$	  & $-5.2 \pm 0.5$	\\
  22 		& $-3.2 \pm 0.4$	& $1.8$	  & $-5.0 \pm 0.4$	\\
  24 		& $-3.8 \pm 0.4$	& $2.0$	  & $-5.8 \pm 0.4$	\\
  26 		& $-3.4 \pm 0.5$	& $2.2$		& $-5.6 \pm 0.5$	\\
  28 		& $-3.1 \pm 0.4$	& $2.3$		& $-5.4 \pm 0.4$	\\
  \hline
  Experimental\cite{morton1995energetic}	& \textendash 		& \textendash	& $-5.2 \pm 0.2$	\\
  \hline
 \end{tabular}
\end{table}

Having validated our approach, we turn our attention to the pathways emerging from the WT-MetaD calculations. These are extracted by visual inspection of the trajectories. We identified eight different pathways (\textbf{A}-\textbf{H} in Figure \ref{fgr:pathways}). Of those eight, five (\textbf{A}-\textbf{E}) have already been reported in the literature\cite{nunes2018escape,rydzewski2018finding} (see Table \ref{tab:pathways}). 

\begin{table}
\caption{Comparison of the different techniques with respective results on the number of identified pathways.}
 \label{tab:pathways}
 \begin{tabular}{ccccc}
  \hline
  Source & CV-based & Force Field   & Water     & \# Pathways \\
  \hline
  Miao et al. \cite{miao2015gaussian}               & No    & Amber99SB & explicit & 1 \\
  Wang et al. \cite{wang2017biomolecular}           & Yes   & CHARMM22* & explicit & 1 \\
  Nunes-Alves et al.\cite{nunes2018escape}          & No    & CHARMM36  & implicit & 4 \\
  Rydzewski et al. \cite{rydzewski2018finding}      & Yes   & OPLS-AA/L & explicit & 5 \\
  \hline
  This work                                         & Yes   & Amber14SB & explicit & 8 \\
  \hline
 \end{tabular}
\end{table}

\begin{figure}[h!]
 \includegraphics[width=0.7\textwidth]{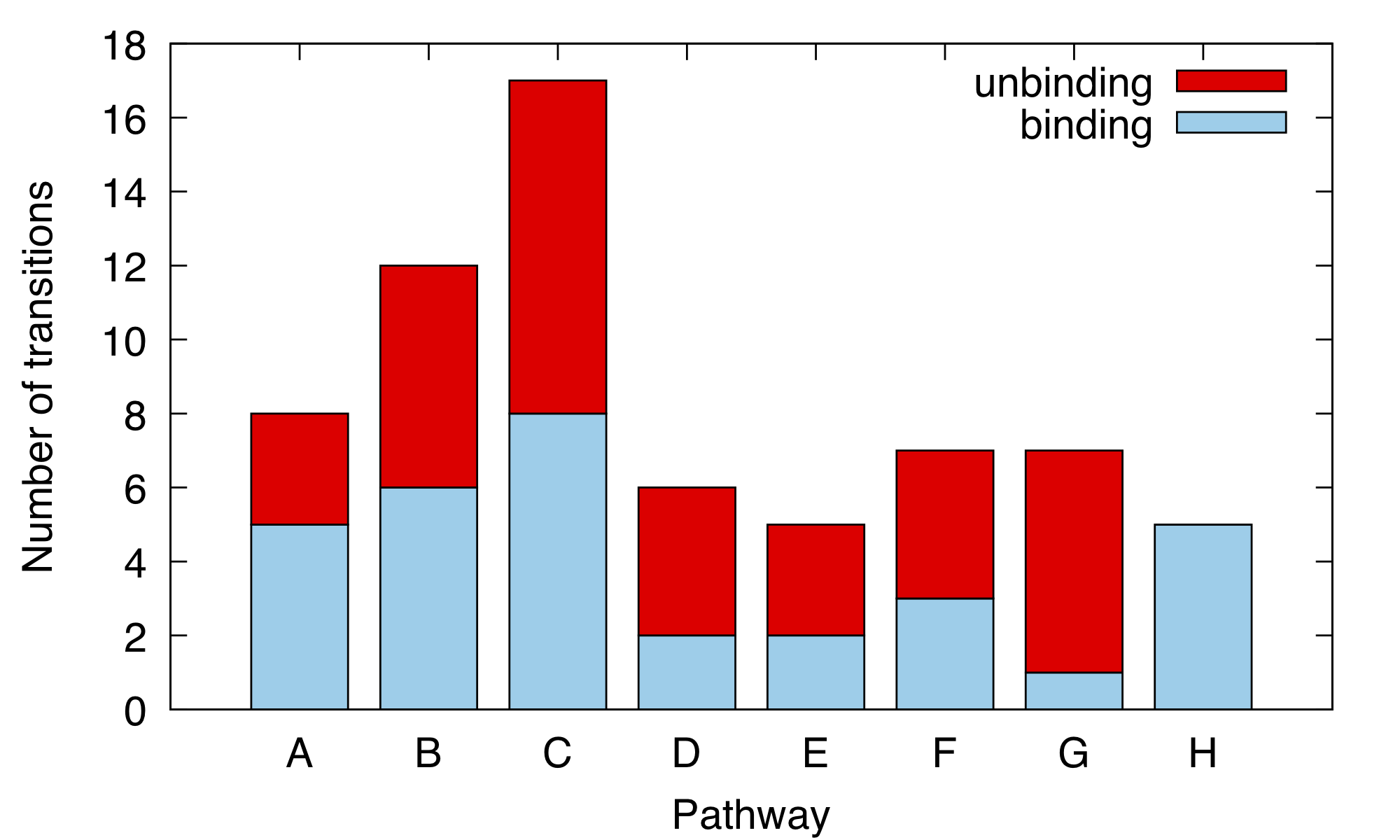}
 \caption{Stacked histogram of the observed binding/unbinding transitions via different pathways. The data is coming from all the 5 metadynamics simulations with different $\rho_{\text{s}}$ values (see equation \ref{eqn:correction}). The unbinding and binding processes are in red and blue, respectively.}
 \label{fgr:histo}
\end{figure}

We discuss here only the new found ones (for a full description of all the 8 pathways, see the SI). In pathway \textbf{F} the ligand moves toward the helices H, I, and J (see Figure \ref{fgr:T4L}). In contrast with pathway \textbf{C}, that shares this first phase, during the solvation the ligand does not interact with the helices G and H, and passes through the helices F and I. Pathway \textbf{G} involve the passage of the ligand between the 3 helices H, I and , slightly broadening the distance between helices H and J. In contrast to all the others, \textbf{H} is the only one-way pathway (see Figure \ref{fgr:histo}): the ligand can enter between the helices E and H, but it cannot pursue the same route for the unbinding transition because from the binding pose VAL149 and MET102 impede its motion toward helices H and E. Interestingly, the single-point mutation M102Q has been studied previous in the literature to favor the binding of polar ligand inside T4L\cite{graves2005decoys,baase2010lessons}. \\

In conclusion, we presented a highly efficient approach to sample all the possible binding/unbinding pathways for a ligand inside a host protein. We shown that the choice of the potential radius $\rho_{\text{s}}$ does not influence the agreement with experimental data of our $\Delta G_{b}^{0}$ estimation after the application of the entropic correction. Our approach was also able to identify all the 5 previously found binding pathway, and to adding 3 new ones.\\
Our technique is totally general and does not need any information regarding the pathways and the bound state. \\

All the data and PLUMED input files required to reproduce the results reported in this paper are available on PLUMED-NEST (www.plumed-nest.org), the public repository of the PLUMED consortium\cite{plumed2019consortium}, as plumID:19.017.

\begin{acknowledgement}
The authors want to thank GiovanniMaria Piccini, Michele Invernizzi, Loris Di Cairano, Luca Maggi, and Mauro Pastore for fruitful discussion.
This project has received funding from the European Union's Horizon 2020 Research and Innovation Programme under Grant Agreement No. 785907 (HBP SGA2).
\end{acknowledgement}

\begin{suppinfo}

Preparation and simulation details, derivation of the entropic correction, convergence of $\Delta G_{b}^{0}$ estimates with different values of $\rho_{\text{s}}$, a schematic representation of the volume bias, all pathways description, and PLUMED input files are available in the Supporting Information.

\end{suppinfo}

\bibliography{sphere}

\end{document}